\documentclass[11pt]{article}

\usepackage{amsmath,amsfonts,amssymb}

 \usepackage[numbers]{natbib}
\usepackage{graphicx}

\usepackage{hyperref}
\hypersetup{
   colorlinks   =  true,
    linkcolor    = cyan,
    citecolor    = red,
     urlcolor	=magenta,     
 }

\newcommand\be{\begin{equation}}
\newcommand\ee{\end{equation}}
\newcommand\bea{\begin{eqnarray}}
\newcommand\eea{\end{eqnarray}}
  
\newcommand\ka{\Lambda}
\newcommand\kk{\omega}

\newcommand\dsum{\oplus_S}

\newcommand\ca{\mathfrak{c}}

\newcommand\nh{\mathfrak{n}}

 \newcommand\ro{\eta }

\def\>#1{{\bf #1}}

\begin{document}

\
  \vskip1cm

\begin{center}
\noindent
 {\LARGE \bf  
The $\kappa$-Newtonian  and  $\kappa$-Carrollian algebras\\[6pt]  
and their noncommutative spacetimes} 
\end{center}

\medskip 
\medskip 
\medskip 
\medskip

\begin{center}

{\sc Angel Ballesteros$^1$, Giulia Gubitosi$^1$, Ivan Gutierrez-Sagredo$^{1,2}$\\[4pt] and Francisco J.~Herranz$^1$}

\medskip

{$^1$Departamento de F\'isica, Universidad de Burgos, 
09001 Burgos, Spain}

{$^2$Departamento de Matem\'aticas y Computaci\'on, Universidad de Burgos, 
09001 Burgos, Spain}

 \medskip
 
e-mail: {\href{mailto:angelb@ubu.es}{angelb@ubu.es}, \href{mailto:giulia.gubitosi@gmail.com}{giulia.gubitosi@gmail.com}, \href{mailto:igsagredo@ubu.es}{igsagredo@ubu.es}, \href{mailto:fjherranz@ubu.es}{fjherranz@ubu.es}}

\end{center}

\medskip

\begin{abstract}
\noindent
We derive the  non-relativistic $c\to\infty$ and ultra-relativistic $c\to 0$ limits of the $\kappa$-deformed   symmetries and corresponding spacetime in (3+1) dimensions, with and without a cosmological constant.  We apply the theory of Lie bialgebra contractions to the Poisson version of the $\kappa$-(A)dS quantum algebra, and  quantize the resulting contracted Poisson--Hopf algebras, thus giving rise to the $\kappa$-deformation of the Newtonian  (Newton--Hooke and Galilei) and Carrollian (Para-Poincar\'e, Para-Euclidean and Carroll) quantum symmetries, including their deformed quadratic Casimir operators. The  corresponding $\kappa$-Newtonian and   $\kappa$-Carrollian  noncommutative spacetimes are also obtained as the non-relativistic and ultra-relativistic limits of the $\kappa$-(A)dS  noncommutative spacetime. These constructions allow us to analyze the  non-trivial interplay between the quantum deformation parameter $\kappa$, the curvature parameter $\eta$ and the speed of light parameter $c$.
  \end{abstract}
\medskip
\medskip


\bigskip


 \newpage


\section{Introduction}

Quantum gravity research usually focusses on the ``relativistic regime'', namely it considers scenarios that at low energies/large distances reduce either to general or special relativity, with finite speed of light. This is for example the case of noncommutative spacetime models \cite{Doplicher:1994zv, MR1994, Madore:2000en, Balachandran:2005eb}. They provide a formalization of the expected fuzzy behaviour of spacetime close to the Planck length $L_{p}\simeq 10^{-35} \text{m}$  \cite{AmelinoCamelia:2008qg,Garay:1994en, Hossenfelder:2012jw}, where quantum-gravitational effects break the usual description of spacetime in terms of a smooth pseudo-Riemannian manifold. These models describe a  ``special relativistic regime'' of quantum gravity, since when the noncommutativity parameter  (related to the Planck length) vanishes, one recovers the usual description of Minkowski spacetime.

Over the last two decades, motivated by the need to make contact with phenomenological studies in the astrophysical and cosmological setting \cite{Rosati:2015pga, Barcaroli:2015eqe}, such models were extended to include a non-vanishing spatial curvature, thus developing noncommutative versions of the (Anti-)de Sitter ((A)dS) models (see~\cite{Brunoc,Stein,BM,VH, Ivetic:2013yga,BRHAdvances,Zoupanos} and references therein). As the understanding of these models advanced, it became increasingly clear that the interplay between the curvature parameter and the Planck-scale noncommutativity parameter is non-trivial \cite{Marciano:2010gq}. For example, some kinds of deformations of a particle dispersion relation might only appear when  the two parameters are both non-zero \cite{Barcaroli:2015eqe,Marciano:2010gq}, and the rotation sector might be deformed \cite{Ballesteros:2017pdw}.

Having realised that the introduction of a curvature parameter might have non-trivial consequences in quantum spacetime models, it is reasonable to wonder what role does the speed of light parameter play in this context. In particular, the question of what residual quantum-gravity effects might survive in the non-relativistic ($c\to \infty$) and ultra-relativistic ($c\to 0$) limits should be investigated.\footnote{Despite the possibly misleading terminology that is commonly used, both the non-relativistic and the ultra-relativistic limits of Poincar\'e invariant models retain invariance under some group of kinematical symmetries, the Galilei and the Carroll group, respectively. This is also true when the same limits are applied to systems  with (Anti-)de Sitter invariance. In this case, the two limits produce models that are invariant under the so-called Newton--Hooke and curved Carroll groups of symmetries, respectively (each of them can in turn have positive or negative cosmological constant).} These are issues with possible phenomenological consequences: the $c\to \infty$ limit is relevant for the search of quantum-gravity signatures in systems where the typical velocities are  small compared to the speed of light, such as in atom interferometry \cite{Garay:1999cy, Ellis:1983jz}, while the $c\to 0$ limit can be related to quantum effects in a strong gravity regime \cite{Dautcourt:1997hb, Bergshoeff:2017btm, PintoNeto:1998mz}.

In previous work \cite{SnyderG}, some of us started to uncover that indeed the non-relativistic and ultra-relativistic limits of quantum spacetime models are non-trivial. This was done by studying these limits in the context of  Snyder noncommutativity with zero spacetime curvature. We found that, contrary to what commonly assumed, the separation of space and time that one would usually expect in the non-relativistic and ultra-relativistic limits is not realised, because of a residual  noncommutativity between time and space coordinates. Moreover, the limits are realised in a non-trivial way, since they are somewhat exchanged when looking at the corresponding momentum space geometry of each model.

In this paper we further investigate these issues in the context of one of the most studied models of quantum spacetime, characterized by the so-called $\kappa$-deformation. The flat case, known as the $\kappa$-Minkowski spacetime with associated $\kappa$-Poincar\'e symmetry, has been widely investigated in the context of quantum gravity research, uncovering a rich structure and a variety of possible phenomenological consequences \cite{KowalskiGlikman:2002we, Bruno:2001mw, Gubitosi:2013rna, AmelinoCamelia:2011nt, AmelinoCamelia:1999pm, Gubitosi:2019ymi}. In more recent times, the model has been extended to include non-vanishing curvature, with the development of the $\kappa$-(A)dS spacetime~\cite{BGH2019spacetime} and its associated $\kappa$-(A)dS symmetries~\cite{BHMN2017kappa3+1}. It was indeed by using these models that the interplay between spacetime curvature and noncommutativity was most precisely characterized. Here we show that studying the non-relativistic and ultra-relativistic limits of these models allows us to investigate the possible interplay between the noncommutativity parameter, the cosmological constant and the speed of light all at once.

We compute for the first time the Poisson version of the $\kappa$-deformed Newtonian and Carrollian quantum algebras, including their extension with non-vanishing curvature. These can be obtained as the $c\to \infty$ and $c\to 0$ contractions of the $\kappa$-(A)dS algebra, respectively.\footnote{The zero-curvature case is trivially obtained by sending the cosmological constant parameter to zero.} Furthermore, we obtain the noncommutative spacetimes that are invariant under such symmetries, which we call the $\kappa$-Newtonian and $\kappa$-Carrollian spacetimes (under these names we include both the vanishing and non-vanishing  cosmological constant cases), showing that they are the 
$c\to \infty$ and $c\to 0$ limits of the $\kappa$-(A)dS spacetime. We find that, similarly to what happens for the Snyder model, also with the $\kappa$-deformation space and time maintain a residual noncommutativity in both the non-relativistic and the ultra-relativistic limits. Moreover, the form of the Casimir of the algebra of symmetries in the non-relativistic limit suggests a residual interplay between spatial momenta and energy (this is not the case in the ultra-relativistic limit). The study of the $\kappa$-deformation with non-zero spatial curvature lets us also appreciate the interplay of all of the three parameters for the model: while the non-relativistic limits preserves the deformation of the algebra of rotations that is induced by the curvature parameter in the $\kappa$-(A)dS algebra, in the ultra-relativistic limit the standard algebra of rotations is recovered, even when the curvature is non-vanishing.

The structure of the paper is as follows. In Section \ref{sec2} we review the classical (A)dS algebra and its non-relativistic and ultra-relativistic limits. These are obtained after an appropriate $c$-dependent rescaling of the symmetry generators. In Section \ref{sec3} we review the contraction procedure that can be applied to quantum algebras to obtain the non-relativistic and ultra-relativistic limits, and use it to compute the $\kappa$-Newtonian algebra with non-vanishing cosmological constant and its $\kappa$-Galilei algebra limit. In Section \ref{sec4} we apply the same procedure to obtain the $\kappa$-Carrollian algebras, with vanishing and non-vanishing cosmological constant. In Section \ref{sec5}  we construct the corresponding quantum spacetimes  that are invariant under the $\kappa$-Newtonian and $\kappa$-Carrollian symmetries, respectively. For these spacetimes we provide both the Poisson brackets structure and their quantization. The final Section \ref{sec6} provides some concluding remarks and outlook.


\section{The non-relativistic and ultra-relativistic limits of (A)dS}\label{sec2}

We start by reviewing the limiting procedures that bring us from the classical (A)dS algebra to its non-relativistic ($c\to \infty$) and ultra-relativistic ($c\to 0$) limits. We show that in order to obtain consistent results, these limits need to be taken after an appropriate rescaling of the symmetry generators. The procedure we expose in this section will be adapted to the $\kappa$-deformed algebra and spacetime in the following sections.

Let us consider the kinematical algebras of isometries of the (3+1)D (A)dS and Minkowskian spacetimes in the basis spanned by the time translation generator $P_0$, space translations $\>P=(P_1,P_2,P_3)$, boosts $\>K=(K_1,K_2,K_3)$ and rotations $\>J=(J_1,J_2,J_3)$.  These three Lorentzian Lie algebras can collectively be  described in terms of the cosmological constant $\ka$ 
as  a one-parameter family of Lie algebras   denoted by AdS$_\ka$ whose commutation relations read
\bea
\begin{array}{lll}
[J_a,J_b]=\epsilon_{abc}J_c ,& \quad [J_a,P_b]=\epsilon_{abc}P_c , &\quad
[J_a,K_b]=\epsilon_{abc}K_c , \\[2pt]
\displaystyle{
  [K_a,P_0]=P_a  } , &\quad\displaystyle{[K_a,P_b]=\delta_{ab} P_0} ,    &\quad\displaystyle{[K_a,K_b]=-\epsilon_{abc} J_c} , 
\\[2pt][P_0,P_a]=-\ka  \,K_a , &\quad   [P_a,P_b]=\ka \,\epsilon_{abc}J_c , &\quad[P_0,J_a]=0  ,
\end{array}
\label{aa}
\eea
where  from now on sum over repeated indices will be  understood   and $a,b,c=1,2,3$. The family  AdS$_\ka$ is endowed with two Casimir 
operators; one of them is the quadratic Casimir coming from the  Killing--Cartan form, namely,
\be
{\cal C}
= P_0^2-\>P^2 +\ka \left(  \>K^2- \>J^2\right) ,
\label{ab}
\ee
while the other one is  fourth-order, and is related to the Pauli--Lubanski 4-vector. Its explicit form can be found in~\cite{Herranz:1997us,Herranz:2006un}.
Hence    AdS$_\ka$   is   the  AdS    algebra $\mathfrak{so}(3,2)$  for  $\ka<0$,   the 
 dS    algebra $\mathfrak{so}(4,1)$  when   $\ka>0$,   and the  Poincar\'e algebra  $\mathfrak{iso}(3,1)$ for  $\ka=0$. This latter case is just the spacetime contraction  of the (A)dS algebras  which is associated with the composition of the   parity $\mathcal {P}$ and   time-reversal $\mathcal{T}$   involutive automorphisms~\cite{BLL}, which acts on the generators as follows:
    \be
  \mathcal{PT}( P_0,\>P,\>K,\>J)=  (-P_0,-\>P,\>K,\>J).
\label{ac}
\ee

The three (3+1)D Lorentzian spacetimes  of constant curvature are obtained as the coset spaces that we will denote as
\be
\>{AdS}^{3+1}_\ka= G/H ,\qquad H= {\rm SO}(3,1)=\langle \>K,\>J\rangle ,
\label{ad}
\ee 
where $G$ is the Lie group with Lie algebra  AdS$_\ka$ and $H$ is the isotropy subgroup, namely the Lorentz subgroup generated by boosts and rotations. When $\ka<0$ we obtain the AdS space, when $\ka<0$ the dS one and the $\ka=0$ coset space is just the Minkowski spacetime. Hence $\>{AdS}^{3+1}_\ka$ encompasses the family of   symmetrical homogeneous Lorentzian spacetimes (with involution (\ref{ac}))   of constant sectional curvature $\kk=-\ka$.


\subsection{The non-relativistic limit: Newtonian algebras and spacetimes}

 In order to apply the non-relativistic limit within  the  family AdS$_\ka$  we first introduce explicitly the speed of light $c$ via the map
\be
\>P\to \frac 1 c\, \>P,\qquad  \>K\to \frac 1 c\, \>K.
\label{add}
\ee
Applying this to the commutation rules  (\ref{aa}) and  taking the limit $c\to \infty$  yields
\bea
\begin{array}{lll}
[J_a,J_b]=\epsilon_{abc}J_c ,& \quad [J_a,P_b]=\epsilon_{abc}P_c , &\quad
[J_a,K_b]=\epsilon_{abc}K_c , \\[2pt]
\displaystyle{
  [K_a,P_0]=P_a  } , &\quad\displaystyle{[K_a,P_b]=0} ,    &\quad\displaystyle{[K_a,K_b]=0} , 
\\[2pt][P_0,P_a]=-\ka  \,K_a , &\quad   [P_a,P_b]=0 , &\quad[P_0,J_a]=0  .
\end{array}
\label{ae}
\eea
The non-relativistic limit of the second-order Casimir  is obtained by transforming (\ref{ab}) through the map (\ref{add}) and then taking the limit $\lim_{c\to\infty}(-{\cal C}/c^2)$, leading to
\be
{\cal C}
= \>P^2 -\ka\, \>K^2 .
\label{af}
\ee
We recall that this procedure is just an In\"on\"u--Wigner contraction~\cite{Inonu:1953sp} which corresponds to    the speed-space contraction associated with  the parity  $\mathcal{P}$ automorphism~\cite{BLL}
  \be
 \mathcal{P}( P_0,\>P,\>K,\>J)=  (P_0,-\>P,-\>K,\>J).
\label{ag}
\ee

In this way we have obtained the family of non-relativistic or Newtonian Lie algebras, that we will denote as  $\nh_\ka$: the expanding Newton--Hooke (NH) algebra $\nh_+$ for $\ka>0$, the oscillating NH algebra $\nh_-$  for $\ka< 0$ and   
the Galilei algebra $\nh_0$ for $\ka=0$~\cite{SnyderG,Herranz:1997us,Herranz:2006un,BLL,BHOS1994global,Aldrovandi,Duval:2014uoa,Figueroa-OFarrill:2017ycu,Gomis:2019}. These algebras have  the following structure (the notation $\dsum$ stands for the semidirect sum):
  \be
\begin{array}{llll}
\nh_+= \mathbb{R}^6\dsum\bigr(  \mathfrak{so}(1,1) \oplus   \mathfrak{so}(3) \bigl),&\  
  \mathbb{R}^6=\langle \>P,\>K\rangle ,&\   \mathfrak{so}(1,1)=\langle   P_0\rangle,&\   \mathfrak{so}(3)= \langle  \>J \rangle.
\\[4pt] 
\nh_-=  \mathbb{R}^6\dsum\bigr(  \mathfrak{so}(2) \oplus   \mathfrak{so}(3) \bigl),&\  
  \mathbb{R}^6=\langle \>P,\>K\rangle ,&\   \mathfrak{so}(2)=\langle   P_0\rangle,&\   \mathfrak{so}(3)= \langle  \>J \rangle.
\\[4pt] 
\nh_0=   \mathbb{R}^4\dsum\bigr(    \mathbb{R}^3 \dsum  \mathfrak{so}(3) \bigl)\equiv  \mathfrak{iiso}(3) ,&\  
   \mathbb{R}^4=\langle P_0, \>P\rangle ,&\     \mathbb{R}^3=\langle  \>K \rangle,&\   \mathfrak{so}(3)= \langle  \>J \rangle.
   \end{array}
\label{ah}
\ee

The three corresponding (3+1)D Newtonian spacetimes  of constant curvature are constructed as the coset spaces
\be
\>{N}^{3+1}_\ka= {\rm N}_\ka/H ,\qquad H= {\rm ISO}(3)=\langle \>K,\>J\rangle ,
\label{ai}
\ee 
where $ {\rm N}_\ka$ is the Lie group with Lie algebra  $\nh_\ka$ and $H$ is the isotropy subgroup  of rotations and (commuting) Newtonian boosts, which is isomorphic to ${\rm ISO}(3)$.  These three non-relativistic spacetimes have the same constant sectional  curvature $\kk=-\ka$ as  their Lorentzian counterparts. We point out that $\kk$  is the curvature of the ``main" metric (which is degenerate and provides the ``absolute-time" description), but there exists an additional invariant foliation under the Newtonian group action with a ``subsidiary" 3D non-degenerate 
  Euclidean spatial metric restricted to each leaf of the foliation (see, e.g.~\cite{SnyderG,conf} for explicit metric models).


\subsection{The ultra-relativistic limit: Carrollian algebras and spacetimes}

The ultra-relativistic limit of the family AdS$_\ka$ can be performed by introducing the speed of light $c$ via  the map~\cite{BLL,LevyLeblondCarroll} 
\be
P_0\to  c\, P_0,\qquad  \>K\to    c\, \>K.
\label{aj}
\ee
Applying this to the commutation relations  (\ref{aa}) and  taking the limit $c\to 0$ generates  the family of Lie algebras 
\bea
\begin{array}{lll}
[J_a,J_b]=\epsilon_{abc}J_c ,& \quad [J_a,P_b]=\epsilon_{abc}P_c , &\quad
[J_a,K_b]=\epsilon_{abc}K_c , \\[2pt]
\displaystyle{
  [K_a,P_0]=0 } , &\quad\displaystyle{[K_a,P_b]=\delta_{ab} P_0} ,    &\quad\displaystyle{[K_a,K_b]= 0} , 
\\[2pt][P_0,P_a]=-\ka  \,K_a , &\quad   [P_a,P_b]=\ka \,\epsilon_{abc}J_c , &\quad[P_0,J_a]=0  .
\end{array}
\label{ak}
\eea
The ultra-relativistic limit of the second-order AdS$_\ka$ Casimir is obtained by transforming (\ref{ab}) under the map   (\ref{aj}) and taking the limit  $\lim_{c\to0}c^2\,{\cal C}$, thus yielding
\be
{\cal C}
=  P_0^2 +\ka\, \>K^2 .
\label{al}
\ee
This  process defines  an In\"on\"u--Wigner contraction~\cite{Inonu:1953sp} which is interpreted as    the speed-time contraction associated with  the time-reversal  $\mathcal{T}$ involution~\cite{BLL}
  \be
 \mathcal{T}( P_0,\>P,\>K,\>J)=  (-P_0,\>P,-\>K,\>J).
\label{am}
\ee

Thus we have obtained the family of Carrollian Lie algebras,   denoted $\ca_\ka$,  which comprises  the   Para-Euclidean   $\ca_+$,  
Para-Poincar\'e   $\ca_-$ and the (proper) Carroll $\ca_0$ algebras~\cite{Bergshoeff:2017btm,SnyderG,BLL,BHOS1994global,Duval:2014uoa, Figueroa-OFarrill:2017ycu,Gomis:2019,LevyLeblondCarroll,Gomis:2014, KowalskiGlikman:2014,Hartong:2015xda,Cardona:2016ytk, Daszkiewicz2019}. Their internal structure can be described  as follows~\cite{SnyderG}: 
  \be
\begin{array}{lll}
\ca_+\equiv \mathfrak{i'so}(4) =  \mathbb{R}'^4 \dsum  \mathfrak{so}(4) ,&\  
  \mathbb{R}'^4=\langle P_0,\>K\rangle ,&\   \mathfrak{so}(4)=\langle   \>P,\>J\rangle .
\\[4pt] 
\ca_- \equiv \mathfrak{i'so}(3,1) =  \mathbb{R}'^4 \dsum  \mathfrak{so}(3,1) ,&\  
  \mathbb{R}'^4=\langle P_0,\>K\rangle ,&\   \mathfrak{so}(3,1)=\langle   \>P,\>J\rangle .
\\[4pt] 
\ca_0\equiv \mathfrak{i'iso}(3) =  \mathbb{R}'^4 \dsum \bigr( \mathbb{R}^3 \dsum  \mathfrak{so}(3)  \bigl)   ,&\  
   \mathbb{R}'^4=\langle P_0, \>K\rangle ,&\     \mathbb{R}^3=\langle  \>P \rangle, \quad\  \mathfrak{so}(3)= \langle  \>J \rangle  .
   \end{array}
\label{an}
\ee
We remark that $\mathfrak{i'so}(4)$ is isomorphic to the Euclidean algebra $\mathfrak{iso}(4)$ and $\mathfrak{i'so}(3,1) $ to the Poincar\'e algebra  $\mathfrak{iso}(3,1)$, although they are physically different algebras. In fact,  the $'$ notation for the Para-Poincar\'e algebra  $ \mathfrak{i'so}(3,1) $    means that $ \mathfrak{so}(3,1)=\langle  \>P,\>J\rangle$ acts on $ \mathbb{R}'^4=\langle P_0,\>K\rangle$ through the contragredient of the vector representation, while in the Poincar\'e algebra  $ \mathfrak{iso}(3,1) $ the Lorentz subalgebra $ \mathfrak{so}(3,1)=\langle  \>K,\>J\rangle $ acts on $ \mathbb{R}^4=\langle P_0,\>P\rangle$ through the   vector representation, and similarly for the two remaining Carrollian algebras (see~\cite{Azcarraga} for details).

The three (3+1)D Carrollian spacetimes  of constant curvature are  identified with the coset spaces
\be
\>{C}^{3+1}_\ka= {\rm C}_\ka/H ,\qquad H= {\rm ISO}(3)=\langle \>K,\>J\rangle ,
\label{a0}
\ee 
where $ {\rm C}_\ka$ is the Lie group with Lie algebra  $\ca_\ka$ and $H$ is again the isotropy subgroup ${\rm ISO}(3)$  spanned by rotations and (commuting) Carrollian boosts. We stress that now such Carrollian spacetimes have    sectional  curvature $\kk=+\ka$, instead of $\kk=-\ka$   as in the Lorentzian and Newtonian spacetimes. In particular, the ``main" metric for Carrollian spacetimes is again degenerate and provides an ``absolute-space" geometry with curvature $\kk=+\ka$. But there does also exist
an invariant foliation preserved by the Carrollian group action which is characterized by a  ``subsidiary" 1D  time metric that is restricted to each leaf of the foliation~\cite{SnyderG}.

For the sake of clarity, in Table~\ref{table1} we summarize the nine kinematical algebras (Lorentzian, Newtonian and Carrollian)  and the corresponding spacetimes.

  
\begin{table}[t]
{\footnotesize
\caption{\small  Lorentzian, Newtonian and Carrollian  algebras  together with their corresponding (3+1)D   homogeneous  spacetimes  with constant sectional curvature $\kk$  according to the value of $\ka$.}
\label{table1}
  \begin{center}
\noindent
\begin{tabular}{l l l  }
\hline
\\[-0.2cm]
\multicolumn{3}{c}{Lorentzian  algebras and   spacetimes} \\[0.2cm]
\hline
\\[-0.2cm] 
 $\bullet$ AdS  &  $\bullet$ Poincar\'e &  $\bullet$ dS  \\[0.2cm]
$\mathfrak{so}(3,2)$: $\ka<0$, $\kk=-\ka>0$  & $\mathfrak{iso}(3,1)$: $\ka=\kk=0$  & $\mathfrak{so}(4,1)$: $\ka>0$,  $\kk=-\ka<0$  \\[0.2cm]
  $\>{AdS}^{3+1}={\rm SO}(3,2)/{\rm SO}(3,1)$  &   $\>{M}^{3+1}={\rm ISO}(3,1)/{\rm SO}(3,1)$ & $\>{dS}^{3+1}={\rm SO}(4,1)/{\rm SO}(3,1)$ \\[0.2cm]
\hline
\\[-0.2cm]
\multicolumn{3}{c}{Newtonian  algebras and spacetimes} \\[0.2cm]
\hline
\\[-0.2cm] 
 $\bullet$ Oscillating   NH&  $\bullet$ Galilei &  $\bullet$ Expanding NH \\[0.2cm]
$\nh_-  =\mathbb{R}^6\dsum\bigr(  \mathfrak{so}(2) \oplus   \mathfrak{so}(3) \bigl)$ & $\nh_0=\mathfrak{iiso}(3)$ &  $\nh_+=\mathbb{R}^6\dsum\bigr(  \mathfrak{so}(1,1) \oplus   \mathfrak{so}(3) \bigl)$     \\[0.2cm]
$\ka<0$, $\kk=-\ka>0$  & $\ka=\kk=0$  &   $\ka>0$, $\kk=-\ka<0$ \\[0.2cm]
  $\>{N}_-^{3+1}={\rm N}_-/{\rm ISO}(3)$   &   $\>{N}_0^{3+1}\equiv \>{G}^{3+1}={\rm IISO}(3)/{\rm ISO}(3)$ &$\>{N}_+^{3+1}={\rm N}_+/{\rm ISO}(3)$  \\[0.2cm]
\hline
\\[-0.2cm]
\multicolumn{3}{c}{Carrollian  algebras and spacetimes} \\[0.2cm]
\hline
\\[-0.2cm] 
 $\bullet$ Para-Poincar\'e  &  $\bullet$  Carroll&  $\bullet$ Para-Euclidean \\[0.2cm]
$\ca_- = \mathfrak{i'so}(3,1) $: $\ka=\kk<0$  & $\ca_0= \mathfrak{i'iso}(3) $: $\ka=\kk=0$  & $\ca_+= \mathfrak{i'so}(4)$: $\ka=\kk>0$   \\[0.2cm]
$\>{C}_-^{3+1}={\rm I'SO(3,1)}/{\rm ISO}(3)$    &   $\>{C}_0^{3+1}\equiv \>{C}^{3+1}={\rm I'ISO}(3)/{\rm ISO}(3)$  &  $\>{C}_+^{3+1}={\rm I'SO(4)}/{\rm ISO}(3)$     \\[0.2cm]
\hline
\end{tabular}
 \end{center}
}
 \end{table} 
 


\section{The   $\kappa$-deformation of the Newtonian  algebras}
\label{sec3}

In the previous section we have seen how to perform the non-relativistic limit of the AdS$_\ka$ algebra with commutation relations (\ref{aa}): one needs to introduce the speed of light parameter via a rescaling of the symmetry generators, eq. \eqref{add}, and then perform the limit $c\to \infty$. When dealing with a quantum algebra, in our case the $\kappa$-deformation of the  AdS$_\ka$ algebra, one needs to identify the appropriate rescaling of the quantum deformation parameter $\kappa$ that, along with the map \eqref{add}, allows us to obtain meaningful expressions in the $c\to \infty$ limit (the idea that in quantum group contractions the deformation parameter has to be transformed was introduced for the first time in~\cite{CGST1991heisemberg, CGST1992}).
In order to do so we start by recalling the fundamental structures that underlie the $\kappa$-AdS$_\ka$ algebra. A more detailed discussion can be found in \cite{BGH2019spacetime,BHMN2017kappa3+1}.

The $\kappa$-deformation of the AdS$_\ka$ algebra  can be generated  by the   classical $r$-matrix  given by~\cite{BGH2019spacetime,BHMN2017kappa3+1}
\be
r_\ka=\frac{1}{\kappa}( K_1 \wedge P_1 + K_2 \wedge P_2 + K_3 \wedge P_3 + \ro\,  J_1 \wedge  J_2),
\label{ba}
\ee
where the parameter    $\ro$ is related to the cosmological constant via
\be
\ro^2:= {-\ka}  .
\label{bb}
\ee
The parameter $\ro$  can be either  real   or  pure imaginary number  for AdS ($r_-$)  and dS ($r_+$), respectively. When $\ro=\ka =0$ we recover the $\kappa$-Poincar\'e $r$-matrix~\cite{Maslanka1993,Zakrzewski1994poincare}
\be
r_0=\frac{1}{\kappa}( K_1 \wedge P_1 + K_2 \wedge P_2 + K_3 \wedge P_3 ) ,
\label{bc}
\ee
underlying the  well known $\kappa$-Poincar\'e algebra and group~\cite{MR1994,LRNT1991,LNR1991realforms,GKMMK1992,LNR1992fieldtheory}.

The $\kappa$-AdS$_\ka$   $r$-matrix (\ref{ba}) is a solution of the modified classical Yang--Baxter equation and leads to a quasi-triangular Lie bialgebra 
through the commutator
\be
\delta(X) = [X \otimes 1 + 1 \otimes X,r_\ka], \qquad \forall X \in   \mbox{AdS$_\ka$},
\label{bd}
\ee
namely~\cite{BGH2019spacetime,BHMN2017kappa3+1},
\begin{align}
\begin{split}
& \delta(P_0)=\delta(J_3)= 0, \qquad \delta(J_1)=\frac \ro\kappa \, J_1 \wedge J_3, \qquad \delta(J_2)= \frac \ro\kappa\,   J_2 \wedge J_3 ,\\
& \delta(P_1)= \frac{1}{\kappa} (P_1 \wedge P_0 - \ro\, P_3 \wedge J_1 - \ro^2 K_2 \wedge J_3 + \ro^2 K_3 \wedge J_2) ,\\
& \delta(P_2)= \frac{1}{\kappa} (P_2 \wedge P_0 - \ro\, P_3 \wedge J_2 + \ro^2 K_1 \wedge J_3 - \ro^2 K_3 \wedge J_1), \\
& \delta(P_3)= \frac{1}{\kappa} (P_3 \wedge P_0 + \ro\, P_1 \wedge J_1 + \ro\, P_2 \wedge J_2 - \ro^2 K_1 \wedge J_2 + \ro^2 K_2 \wedge J_1), \\
& \delta(K_1)= \frac{1}{\kappa} (K_1  \wedge P_0  + P_2 \wedge J_3 - P_3 \wedge J_2 - \ro\, K_3 \wedge J_1) ,\\
& \delta(K_2)= \frac{1}{\kappa} ( K_2 \wedge P_0  - P_1 \wedge J_3 + P_3 \wedge J_1 - \ro\, K_3 \wedge J_2) ,\\
& \delta(K_3)= \frac{1}{\kappa} ( K_3 \wedge P_0  + P_1 \wedge J_2 - P_2 \wedge J_1 + \ro\, K_1 \wedge J_1 + \ro\, K_2 \wedge J_2).
\label{be}
\end{split}
\end{align} 
Starting from these cocommutators, the complete Poisson analogue of the  $\kappa$-AdS$_\ka$ quantum algebra (including fully explicit expressions for both coproducts and deformed Poisson brackets) was constructed in~\cite{BHMN2017kappa3+1}, and written in terms of    the curvature $\kk\equiv \ro^2$ and quantum deformation parameter $z=\kappa^{-1}$.
Recall that  the term $\ro \, J_1 \wedge  J_2$ in $r_\ka$ (\ref{ba}) gives rise to a sub-Lie bialgebra structure within (\ref{be}) coming from the  Lie subalgebra  $\mathfrak{su}(2)\simeq \mathfrak{so}(3)$  spanned   by the rotation generators, which disappears  in the Poincar\'e case with $\ro=0$. So one of the peculiar effects of the interplay between the curvature and quantum deformation parameters is that the algebra of rotations gets deformed, see also the discussion in \cite{Ballesteros:2017pdw}.

The non-relativistic limit of the $\kappa$-AdS$_\ka$ symmetry can be obtained by applying the Lie bialgebra contraction (LBC) approach introduced in~\cite{BGHOS1995quasiorthogonal}. This completely general procedure starts from a given Lie algebra contraction and studies which transformation of the quantum deformation parameter has to be considered in order to obtain a well-defined and non-trivial Lie bialgebra structure after the contraction is performed. The initial Lie algebra contraction together with the transformation law of the quantum deformation parameter in terms of the contraction parameter defines a specific LBC, which suffices in order to define the appropriate contraction of the full quantum algebra and, through suitable consistency conditions arising from duality relations, the associated contraction of the Poisson--Lie and quantum groups (see~\cite{BGHOS1995quasiorthogonal} for details). 

It is important to stress that in this approach the $r$-matrix and its associated cocommutator $\delta$ given by~\eqref{bd} can behave differently under a given LBC, and the former could diverge while the latter is well-defined.  This is quite natural since for non-semisimple Lie algebras (like the Newtonian and Carrollian ones) there do exist non-coboundary Lie bialgebra structures $\delta$ for which no $r$-matrix can be found, and for  some of these cases they can be obtained as a LBC of a coboundary Lie bialgebra. The LBC that guarantees the existence of a non-vanishing cocommutator under contraction is called a fundamental LBC, while the one that guarantees the existence of a non-vanishing $r$-matrix is called a coboundary one. Indeed, there could also exist some LBC which is simultaneously fundamental and coboundary, and in that case the contracted $\delta$ can consistently be  obtained from the contracted $r$ through the coboundary relation (\ref{bd}). As we will see in the sequel, this analysis will be essential for the obtention of Newtonian and Carrollian Lie bialgebras as contractions from AdS$_\ka$.

As a first step in the computation of the non-relativistic limit of the $\kappa$-AdS$_\ka$ Lie bialgebra, 
we check whether there exists a LBC that is coboundary. To this aim, we transform the $r$-matrix~(\ref{ba}) through the map (\ref{add}),  finding 
\be
r_\ka=\frac{c^2}{\kappa}\left( K_1 \wedge P_1 + K_2 \wedge P_2 + K_3 \wedge P_3  + \frac 1{c^2}\,\ro\,  J_1 \wedge  J_2 \right).
\label{bf}
\ee
Now, the existence of a convergent and non-trivial  $c\to \infty $ limit of $r_\ka$ in (\ref{bf}) implies that the unique solution for the transformation law of the deformation parameter is
\be
\kappa \to c^{-2}\kappa \, .
\label{bg}
\ee
Thus the LBC defined by the Lie algebra contraction~\eqref{add} together with the transformation law~\eqref{bg} gives rise to
\be
r_\ka=\frac{1}{\kappa} ( K_1 \wedge P_1 + K_2 \wedge P_2 + K_3 \wedge P_3     ),
\label{bh}
\ee
which is a common $r$-matrix for the three Newtonian algebras $\nh_\ka$ with commutation rules (\ref{ae}). Nevertheless, if we compute the associated cocommutator $\delta$ through  the coboundary relation (\ref{bd}) we get a trivial result, that is, $\delta(X)=0,$ $\forall X\in \nh_\ka$. Indeed, the same vanishing cocommutator is obtained if the same  maps~\eqref{add} and~\eqref{bg}  are applied onto the cocommutator (\ref{be}) and the limit $c\to \infty$ is computed. Therefore, we conclude that the coboundary LBC associated to the non-relativistic limit of the $\kappa$-deformation does exists but  leads to a trivial structure. As it can be straightforwardly checked, when this LBC is applied to the full $\kappa$-AdS$_\ka$ Poisson--Hopf algebra we get a primitive coproduct and non-deformed commutation rules. Moreover, the Sklyanin bracket generated by the $r$-matrix~\eqref{bh} onto the three Newtonian groups vanishes identically, and therefore the spacetime is commutative.

As we mentioned above, one can also look for a  fundamental LBC which is not coboundary and which gives rise to a non-vanishing  cocommutator in the Newtonian limit. In case this exists,  it should be different from the previous coboundary LBC. By introducing the map (\ref{add}) within the cocommutator (\ref{be}) it is straightforward to check that the transformed $\delta$ does not contain $c$ and, therefore, there is no need to transform the deformation parameter under contraction in order to obtain a non-trivial Newtonian Lie bialgebra. Hence  the non-relativistic contracted cocommutator coincides with the $\kappa$-AdS$_\ka$ one given by (\ref{be}). Note that this new fundamental LBC in which the deformation parameter does not change leads to a divergence for the $c\to\infty$ limit of the $r$-matrix~\eqref{bf}, which is consistent with the fact that the Newtonian Lie bialgebra defined by the cocommutator (\ref{be}) together with the commutation rules~\eqref{ae} is not a coboundary Lie bialgebra. Summarizing, a non-trivial $\kappa$-deformation of the Newtonian algebras can be obtained as a fundamental (and non-coboundary) LBC in which the deformation parameter is not affected by the non-relativistic limit.
 
Therefore, if we apply the map (\ref{add}) to the complete $\kappa$-AdS$_\ka$ Poisson--Hopf algebra given in~\cite{BHMN2017kappa3+1} and perform the limit $c\to \infty$ we get fully convergent expressions for the three $\kappa$-Newtonian algebras. The coproduct $\Delta$ given in~\cite{BHMN2017kappa3+1} is found to be  invariant under such contraction, so we omit it for the sake of brevity, while the contracted Poisson brackets turn out to be
\be
\begin{array}{lll}
\multicolumn{3}{l}
{\displaystyle {\left\{ J_1,J_2 \right\}= \frac{ {\rm e}^{2\ro J_3/\kappa}-1}{2 \ro/\kappa } - \frac{ \ro}{2\kappa} \left(J_1^2+J_2^2\right)  ,\qquad
\left\{ J_1,J_3 \right\}=-J_2 ,\qquad
\left\{ J_2,J_3 \right\}=J_1 , } } \\[10pt]
\left\{ J_1,P_1 \right\}=\frac \ro\kappa   J_1 P_2,  & \quad \left\{ J_1,P_2 \right\}=  P_3-\frac \ro\kappa J_1 P_1   ,  & \quad \left\{J_1, P_3 \right\}=  -P_2  ,  \\[4pt]
 \left\{ J_2,P_1 \right\}=-P_3+\frac \ro\kappa J_2 P_2,  & \quad   \left\{ J_2,P_2 \right\}=  -  \frac \ro\kappa J_2 P_1 ,  & \quad   \left\{ J_2,P_3 \right\}=P_1    ,  \\[4pt]
  \left\{J_3, P_1 \right\}= P_2,  & \quad \left\{ J_3,P_2 \right\}=-P_1 ,  & \quad \left\{ J_3,P_3 \right\}=  0 ,  \\[4pt]
\left\{ J_1,K_1 \right\}=\frac \ro\kappa J_1 K_2 ,  & \quad \left\{ J_1,K_2 \right\}=  K_3 -\frac \ro\kappa   J_1 K_1   ,  & \quad \left\{J_1, K_3 \right\}=  -K_2  ,  \\[4pt]
 \left\{ J_2,K_1 \right\}= -K_3+\frac \ro\kappa J_2 K_2,  & \quad   \left\{ J_2,K_2 \right\}= -   \frac \ro\kappa  J_2 K_1  ,  & \quad   \left\{ J_2,K_3 \right\}=K_1    ,  \\[4pt]
  \left\{J_3, K_1 \right\}= K_2 ,  & \quad \left\{ J_3,K_2 \right\}=-K_1 ,  & \quad \left\{ J_3,K_3 \right\}=  0 ,  \\[4pt]
 \left\{ K_{a}, P_0 \right\}=P_a   ,  & \quad \left\{ P_0, P_a \right\}=\ro^2 K_a ,  & \quad \left\{ P_0,J_a \right\}=  0 ,   
 \label{bi}
 \end{array}
\ee
 \begin{eqnarray*}
 \left\{ K_1,P_1 \right\} \!\!\!&=&\!\!\!   \frac{1}{2\kappa} \left( P_2^2+P_3^2-P_1^2\right) +\frac{\ro^2}{2\kappa}  \left(  K_2^2+K_3^2-K_1^2  \right) ,\\
\left\{ K_2,P_2 \right\}\!\!\!&=&\!\!\!  \frac{1}{2\kappa} \left(P_1^2+P_3^2 -P_2^2\right) +\frac{\ro^2}{2\kappa} \left(  K_1^2+K_3^2 -K_2^2  \right) ,\\
\left\{ K_3,P_3 \right\}\!\!\!&=&\!\!\!  \frac{1}{2\kappa} \left[(P_1+ \ro K_2)^2 +(P_2- \ro  K_1)^2-P_3^2  -\ro^2 K_3^2   \right]  ,
 \end{eqnarray*}
 \begin{eqnarray*}
 \left\{ P_1, K_2 \right\} \!\!\!&=&\!\!\! \frac{1}{\kappa}  \left( P_1 P_2+\ro^2 K_1 K_2 - \ro P_3 K_3   \right) , \\
\left\{ P_2, K_1 \right\}\!\!\!&=&\!\!\!  \frac{1}{\kappa} \left(P_1 P_2 + \ro^2 K_1 K_2+  \ro P_3 K_3  \right)  ,\\
\left\{ P_1, K_3 \right\}\!\!\!&=&\!\!\!   \frac{1}{\kappa} \left(  P_1 P_3 +\ro^2 K_1 K_3+ \ro K_2 P_3\right) ,\\
\left\{ P_3, K_1 \right\}\!\!\!&=&\!\!\!  \frac{1}{\kappa} \left(P_1 P_3+\ro^2 K_1 K_3 - \ro  P_2 K_3 \right) ,\\
\left\{ P_2, K_3 \right\}\!\!\!&=&\!\!\!   \frac{1}{\kappa} \left(P_2 P_3+\ro^2 K_2 K_3  -\ro K_1 P_3 \right) , \\
\left\{ P_3, K_2 \right\}\!\!\!&=&\!\!\!   \frac{1}{\kappa} \left(P_2 P_3+\ro^2 K_2 K_3 +   \ro P_1 K_3 \right) ,\\
\left\{ K_a,K_b \right\} \!\!\!&=&\!\!\!  -\frac{\ro}{\kappa}\, \epsilon_{abc} K_cK_3      ,\qquad   \left\{ P_a,P_b \right\} =-\frac{\ro}{\kappa} \,  \epsilon_{abc} P_cP_3  .
 \end{eqnarray*}
 As usual, the limit $\kappa\to \infty$ corresponds to the non-deformed (``classical") limit with commutation rules (\ref{ae}), expressed as Poisson brackets together with a primitive coproduct. Note that this   $c\to\infty$ limit preserves the non-trivial properties of the rotation sector that were already present in the $\kappa$-AdS$_\ka$ case and is due to the interplay between the curvature and quantum deformation parameters.
 
The deformed version of the second-order Newtonian Casimir   (\ref{af})  is obtained as the limit   $\lim_{c\to\infty}{(-\cal C_\kappa}/c^2)$ after  the $\kappa$-AdS$_\ka$ Casimir ${\cal C_\kappa}$ has been transformed under the automorphism (\ref{add}), yielding
\bea
{\cal C}_\kappa \!\!\!&=&\!\!\!  
  {\rm e}^{P_0/\kappa} \left( \mathbf{P}^2 +\ro^2 \mathbf{K}^2 \right)   \left[ \cosh(\ro J_3/\kappa)+ \frac { \ro^2}{2\kappa^2} (J_1^2+J_2^2){\rm  e}^{-\ro  J_3/\kappa} \right]\label{bj}\\
&&-2\ro^2 {\rm e}^{P_0/\kappa}  \left[ \frac{\sinh(\ro J_3/\kappa)}{ \ro} \, R_3+ \frac 1\kappa \left( J_1 R_1 +J_2  R_2+  \frac { \ro}{2\kappa} (J_1^2+J_2^2)  R_3 \right)  e^{- \ro J_3/\kappa} \right],
\nonumber
\eea
where $ R_a=\epsilon_{abc} K_b P_c$.  By comparison with the classical Newtonian Casimir, we see that the presence of the time translation generator is a purely quantum effect, while the presence of the rotation generators is due to the interplay between the curvature and quantum deformation parameters.  This can be confirmed by looking at the $\kappa$-Galilei Casimir that is written explicitly below, eq.~\eqref{bm}. The Newtonian limit of the deformed fourth-order ``Pauli--Lubanski'' Casimir  presented in~\cite{BHMN2017kappa3+1} can be obtained in a similar manner and leads to a very involved expression.

It is worth stressing that the above results provide a  $\kappa$-Newtonian Poisson--Hopf  algebra   whose
 quantization should  be further performed in order to obtain the noncommutative quantum algebra. In the Newton-Hooke cases (i.e. when $\eta\neq 0$) this task implies taking into account all ordering ambiguities appearing in both the coproduct and  the Poisson brackets and we omit the final expressions for the sake of brevity, since
 the essential structure of the deformation can be fully appreciated at the Poisson bracket level.
 However, for the Galilean case (i.e. with $\ro \to 0$) no ordering ambiguities arise and we get the $\kappa$-Galilei quantum algebra:
 \begin{eqnarray}
&& \Delta ( P_0 ) = P_0 \otimes 1+1 \otimes P_0 ,\nonumber\\
&&  \Delta ( J_a ) =   J_a \otimes 1+1 \otimes J_a , \nonumber\\
&& \Delta ( P_a ) =  P_a \otimes 1+{\rm e}^{-  P_0/\kappa}  \otimes P_a  , \label{bk} \\
&& \Delta ( K_a ) =  K_a \otimes  1+{\rm e}^{-  P_0/\kappa} \otimes K_a+\frac 1\kappa \, \epsilon_{abc}   P_b \otimes J_c  ,
\nonumber
\end{eqnarray}
\be
\begin{array}{lll}
[  J_a,J_b] =\epsilon_{abc}J_c ,& \quad [   J_a,P_b] =\epsilon_{abc}P_c , &\quad
[   J_a,K_b] =\epsilon_{abc}K_c , \\[2pt]
\displaystyle{
  [  K_a,P_0] =P_a  } , &\quad\displaystyle{  [  K_a,K_b] =0  } ,    &\quad
  {\displaystyle  {[  K_a, P_b ]  = \delta_{ab} \, \frac{1}{2\kappa}\, \>P^2  - \frac 1\kappa\, P_a P_b  }} ,
\\[4pt]
[  P_0,P_a] = 0 , &\quad   [  P_a,P_b] =0 , &\quad  
\displaystyle{   [  P_0,J_a] =0  } , 
 \end{array}
\label{bl} 
\ee
  \be
{\cal C}_\kappa=
  {\rm e}^{P_0/\kappa} \, \mathbf{P}^2   .
\label{bm} 
\ee
Note that  the resulting Hopf algebra is written in a bicrossproduct basis~\cite{MR1994,Majida,Majidb,Azcarragab} with respect to the four commuting  translations $ \mathbb{R}^4=\langle P_0,\>P\rangle$. We recall 
that the  $\kappa$-Galilei   algebra was firstly obtained in~\cite{GKMMK1992} through contraction from the $\kappa$-Poincar\'e algebra (but without an  LBC analysis) and was expressed in a ``symmetrical" basis which is related to the bicrossproduct one here considered through a nonlinear map. To the best of our knowledge, the $\kappa$-deformation of the oscillating and expanding Newton-Hooke  algebras $(\ro\ne 0)$ is a completely new result.


\section{The   $\kappa$-deformation of the Carrollian  algebras}
\label{sec4}

The ultra-relativistic limit $c\to 0$ of the $\kappa$-AdS$_\ka$ algebra can be performed by following the same LBC approach~\cite{BGHOS1995quasiorthogonal} described in the previous section. In this case the  Lie algebra contraction map is given by (\ref{aj}), which in the classical case leads to the Carrollian Lie algebra $\ca_\ka$ (\ref{ak}). As done in the previous section, we first check whether there is a coboundary LBC. To this aim we apply the contraction map (\ref{aj}) to the $r$-matrix (\ref{ba}), obtaining
\be
r_\ka=\frac{1}{c\,\kappa}( K_1 \wedge P_1 + K_2 \wedge P_2 + K_3 \wedge P_3 + c\,\ro\,  J_1 \wedge  J_2).
\label{ca}
\ee
Asking that this $r$-matrix has a  well-defined and non-trivial limit for $c\to 0$   implies that the deformation parameter must be transformed as
\be
\kappa\to c\,\kappa .
\label{cb}
\ee
Then the $c\to 0$ limit for the $r$-matrix reads
\be
r_\ka=\frac{1}{ \kappa}( K_1 \wedge P_1 + K_2 \wedge P_2 + K_3 \wedge P_3  ),
\label{cc}
\ee
which again coincides with the  $\kappa$-Poincar\'e $r$-matrix (\ref{bc}) for any value of the cosmological constant parameter $\ro$. The remarkable point now is that, in contrast with the Newtonian case, the cocommutator obtained through the relation (\ref{bd}) is a non-trivial one, namely
\begin{align}
\begin{split}
& \delta(P_0)=\delta(J_a)= 0, \qquad \delta(K_a)=\frac 1\kappa\, K_a\wedge P_0    ,\\
& \delta(P_a)= \frac{1}{\kappa} (P_a \wedge P_0 - \ro^2\, \epsilon_{abc}K_b \wedge J_c)  .
\label{cd}
\end{split}
\end{align} 
Furthermore, if we apply the LBC to the AdS$_\ka$  commutator (\ref{be}) given by the map~\eqref{aj} we find that the very same transformation (\ref{cb}) is the one ensuring the convergence of the cocommutator and gives rise to the same contracted expressions (\ref{cd}). Therefore the LBC defined by the maps (\ref{aj}) and (\ref{cb}) together with the limit $c\to 0$ is both a coboundary and a fundamental one. From it, the ultra-relativistic limit of the coproduct and Poisson brackets can directly be  computed from the    $\kappa$-AdS$_\ka$ Poisson algebra~\cite{BHMN2017kappa3+1} thus giving rise to the 
$\kappa$-Carrollian Poisson--Hopf algebras:
 \begin{eqnarray}
&& \Delta ( P_0 ) = P_0 \otimes 1+1 \otimes P_0 ,\qquad
  \Delta ( J_a ) =   J_a \otimes 1+1 \otimes J_a , \nonumber\\
&& \Delta ( P_a ) =  P_a \otimes 1+{\rm e}^{-  P_0/\kappa}  \otimes P_a-\frac {\ro^2}\kappa \, \epsilon_{abc}   K_b \otimes J_c  , \label{ce} \\
&& \Delta ( K_a ) =  K_a \otimes  1+{\rm e}^{-  P_0/\kappa} \otimes K_a ,
\nonumber
\end{eqnarray}
\bea
\begin{array}{lll}
\left\{J_a,J_b\right\}=\epsilon_{abc}J_c ,& \qquad \left\{J_a,P_b\right\}=\epsilon_{abc}P_c , &\quad
\left\{J_a,K_b\right\}=\epsilon_{abc}K_c , \\[3pt]
\displaystyle{
  \left\{K_a,P_0\right\}=0 } , &\qquad  \displaystyle{\left\{K_a,K_b\right\}= 0} ,    &\quad \left\{P_0,J_a\right\}=0 , 
\\[3pt]
\left\{P_0,P_a\right\}=\ro^2  K_a , &\qquad   \left\{P_a,P_b\right\}=-\ro^2  \epsilon_{abc}J_c , &\quad \\[6pt]
\multicolumn{3}{l}  { \displaystyle{ \left\{K_a,P_b\right\}=\delta_{ab}\left(   \frac{1-{\rm e}^{-2 P_0/\kappa}}{2/\kappa} +\frac{\ro^2}{2\kappa} \,\>K^2  \right)  -\frac{\ro^2}{\kappa}\, K_a K_b . }}
\end{array}
\label{cf}
\eea
Finally, the ultra-relativistic limit of the $\kappa$-AdS$_\ka$ deformed quadratic Casimir  ${\cal C_\kappa}$ is obtained by applying the LBC map onto ${\cal C_\kappa}$ and then computing $\lim_{c\to 0}c^2\,{\cal C_\kappa}$. The final result is 
\be
{\cal C_\kappa}=2\kappa^2\left( \cosh(P_0/\kappa)-1\right) -\ro^2 {\rm e}^{P_0/\kappa}\,\>K^2.
\label{cg}
\ee

By inspection of the expressions above, we notice that now the  rotation sector is trivial, so that the ultra-relativistic limit  erases this kind of  effect of the interplay between the curvature and quantum deformations, contrary to what happens in the non-relativistic limit. Moreover, the only residual mixing between the two parameters is of the form $\eta^{2}/\kappa$ (to the first-order in $1/\kappa$), while in the non-relativistic case we had effects already at the first-order in $\eta$, going as $\eta/\kappa$. This gives a first characterization of the  fact that also the speed of light parameter $c$ has a non-trivial interplay with the curvature and quantum deformation parameters.

The proper $\kappa$-Carroll algebra arises under the limit $\ro\to 0$ and is characterized by the following coproducts, Poisson brackets and Casimir functions:
 \begin{eqnarray}
&& \Delta ( P_0 ) = P_0 \otimes 1+1 \otimes P_0 ,\qquad
  \Delta ( J_a ) =   J_a \otimes 1+1 \otimes J_a , \nonumber\\
&& \Delta ( P_a ) =  P_a \otimes 1+{\rm e}^{-  P_0/\kappa}  \otimes P_a   , \qquad
 \Delta ( K_a ) =  K_a \otimes  1+{\rm e}^{-  P_0/\kappa} \otimes K_a ,\label{ch}
\nonumber
\end{eqnarray}
\bea
\begin{array}{lll}
\left\{J_a,J_b\right\}=\epsilon_{abc}J_c ,& \quad \left\{J_a,P_b\right\}=\epsilon_{abc}P_c , &\quad
\left\{J_a,K_b\right\}=\epsilon_{abc}K_c , \\[3pt]
\displaystyle{
  \left\{K_a,P_0\right\}=0 } , &\quad      { \displaystyle{ \left\{K_a,P_b\right\}=\delta_{ab} \,\frac{1-{\rm e}^{-2 P_0/\kappa}}{2/\kappa}    ,}} &\quad   \displaystyle{\left\{K_a,K_b\right\}= 0} , 
\\[3pt]
\left\{P_0,P_a\right\}=0 , &\quad   \left\{P_a,P_b\right\}=0 , &\quad   \left\{P_0,J_a\right\}=0 ,\end{array}
\label{ci}
\eea
\be
{\cal C_\kappa}=2\kappa^2\left( \cosh(P_0/\kappa)-1\right) .
\label{cj}
\ee

We stress that the full  quantization providing the  three $\kappa$-Carrollian quantum algebras is immediate since the   time translation $P_0$ and the Carrollian boosts do commute among themselves for any value of $\ro$,  so no ordering problems appear and the Poisson brackets can be replaced by commutators. We also remark that the resulting quantum algebras are endowed with a bicrossproduct structure~\cite{MR1994,Majida,Majidb,Azcarragab} but now with respect to $ \mathbb{R}'^4=\langle P_0,\>K\rangle$ (see (\ref{an})).
Finally, we recall that twist deformations of the Carroll algebra have recently been constructed in~\cite{Daszkiewicz2019}. They are different from the deformations constructed in this work  since here the time translation generator $P_0$ plays a prominent role.


\section{The   $\kappa$-Newtonian and  $\kappa$-Carrollian   noncommutative\\ spacetimes}
\label{sec5}
 
In the previous sections we have obtained the $\kappa$-Newtonian and $\kappa$-Carrollian algebras.  Now we will deduce their associated noncommutative spacetimes as the corresponding non-relativistic and ultra-relativistic contractions of the noncommutative $\kappa$-AdS$_\ka$ spacetime, which has been recently constructed in~\cite{BGH2019spacetime}. As we will see in the sequel, the LBC approach will also provide all the tools we need to this aim.

Let us  recall that the  $\kappa$-AdS$_\ka$  Poisson homogeneous spacetime (PHS)  was obtained in~\cite{BGH2019spacetime} through the Sklyanin bracket coming from the $r$-matrix (\ref{ba}) by restricting it to the Poisson subalgebra generated by the local spacetime coordinates   $x^0$ and   $x^a$, which are the group parameters of the time translation $P_0$ and space translations $P_a$, respectively. The resulting  semiclassical $\kappa$-AdS$_\ka$ spacetime is given by the following Poisson brackets
\begin{align}
\begin{split}
\label{da}
&\{x^1,x^0\} =\frac{1}{\kappa}\, \frac{\tanh (\ro x^1)}{\ro \cosh^2(\ro x^2) \cosh^2(\ro x^3)} ,\\
&\{x^2,x^0\} =\frac{1}{\kappa}\,\frac{\tanh (\ro x^2)}{\ro \cosh^2(\ro x^3)} ,\\
&\{x^3,x^0\} =\frac{1}{\kappa}\,\frac{\tanh (\ro x^3)}{\ro},
\end{split}
\end{align} 
\begin{align}
\begin{split}
\label{db}
&\{x^1,x^2\} =-\frac{1}{\kappa}\,\frac{\cosh (\ro x^1) \tanh ^2(\ro x^3)}{\ro} ,\\
&\{x^1,x^3\} =\frac{1}{\kappa}\,\frac{\cosh (\ro x^1) \tanh (\ro x^2) \tanh (\ro x^3)}{\ro} ,\\
&\{x^2,x^3\} =-\frac{1}{\kappa}\,\frac{\sinh (\ro x^1) \tanh (\ro x^3)}{\ro} \, ,
\end{split}
\end{align} 
which is a nonlinear deformation of the $\kappa$-Minkowski spacetime, which is recovered as the
$\ro\to 0$ limit of the previous expressions, namely
\be
\{x^a,x^0\} =\frac{1}{\kappa}\, x^a ,\qquad \{x^a,x^b\} = 0.
\label{dc}
\ee
We stress that since the Poisson brackets among the space sector (\ref{db}) do not vanish when $\ro\ne 0$, the quantization of the $\kappa$-AdS$_\ka$ PHS has to be performed by fixing a precise ordering. As it was shown in~\cite{BGH2019spacetime}, a suitable ordering allows  the quantum $\kappa$-AdS$_\ka$ noncommutative spacetime to be expressed as a  homogeneous quadratic algebra, provided that ambient coordinates are considered.

Let us now focus on the obtention of the $\kappa$-Newtonian noncommutative spacetimes, whose associated Poisson--Hopf algebras were obtained in Section~\ref{sec3}. Since these are non-coboundary deformations with no underlying classical $r$-matrix, the use of the Sklyanin bracket in order to get the noncommutative spacetimes is  precluded. Nevertheless the corresponding PHS can be deduced as the non-relativistic limit of the  $\kappa$-AdS$_\ka$  one, since it is well-known that the clue for performing the contraction at the group level is to impose that the pairing between local coordinates and algebra generators has to be preserved (in other words, products $x^0P_0$ and $x^aP_a$ must be invariant under the contraction mapping, see~\cite{BHOS1995jmp} for details). In this way the contraction  map (\ref{add}) induces the following transformation law for the Newtonian coordinates
\be
x^0\to x^0,\qquad x^a\to c \, x^a,
\label{dd}
\ee
and the non-relativistic limit of the Poisson brackets  (\ref{da}) and (\ref{db}) can be obtained by applying (\ref{dd}) onto them, and recalling that in this fundamental LBC the quantum deformation parameter $\kappa$ is not affected by the contraction. In this way, the $c\to\infty$ limit of the transformed Poisson brackets leads to the $\kappa$-Newtonian Poisson homogeneous  spacetimes:
 \be
 \{x^a,x^0\} =\frac{1}{\kappa}\, x^a ,\quad\  \{x^1,x^2\} =-\frac{\ro}{\kappa}\,   (x^3)^2    ,\quad\  \{x^1,x^3\} =\frac{\ro}{\kappa}\,  x^2 x^3
 ,\quad\  \{x^2,x^3\} =-\frac{\ro}{\kappa}\,  x^1 x^3 .
 \label{ddd}
 \ee
 Note that~\eqref{ddd} are identical to the first-order in $\ro$ of the full $\kappa$-AdS$_\ka$ expressions  (\ref{da})--(\ref{db}), which means that the non-relativistic contraction eliminates all higher order contributions in the cosmological constant parameter. Furthermore, the non-relativistic limit does not produce a complete separation between the space and time coordinates, as happens in the classical case. Here, quantum effects produce a residual mixing in the form of non-trivial brackets between the  time and space sectors of spacetime. This is something that was already noticed in the Snyder noncommutative spacetime model \cite{SnyderG}. Furthermore, the joint effect of the quantum deformation and curvature parameters is still visible in the spatial coordinates brackets. Further comments on this are below.
 
Now, by resorting to~\cite{BGH2019spacetime} we realize that this quadratic algebra can  straightforwardly be quantized by considering the ordered monomials $(\hat x^1)^l\,(\hat x^3)^m\,(\hat x^2)^n$, and the quantum $\kappa$-Newtonian noncommutative spacetimes are thus given by the commutation rules
 \be
 [\hat x^a,\hat x^0] =\frac{1}{\kappa}\, \hat x^a,\qquad 
[\hat x^1,\hat x^2] =-\,\frac{\ro}{\kappa}\,(\hat x^3)^2, \qquad
[\hat x^1,\hat x^3] =\,\frac{\ro}{\kappa}\,\hat x^3 \hat x^2,\qquad
[\hat x^2,\hat x^3] =-\,\frac{\ro}{\kappa}\,\hat x^1\hat x^3 ,
\label{de}
\ee
which define a noncommutative and associative  (Jacobi identities hold) homogeneous quadratic algebra. 

We stress that the time-space commutators $ [\hat x^a,\hat x^0] $ in~\eqref{de} are  the same as in the $\kappa$-Minkowski spacetime $\>{M}^{3+1}_\kappa$ (\ref{dc}), but the noncommutative space coordinates define a quadratic subalgebra which,  in the same manner as in the $\kappa$-AdS$_\ka$ case, can be identified~\cite{Grabowski1990brno,Grabowski1995} with a subalgebra of Woronowicz's quantum ${\rm SU}(2)$ group~\cite{Woronowicz1987tsu2,Woronowicz1987cmp,VaksmanSoibelman1988,ChaichianDemichev1996book}. In fact, there does exist a Casimir operator for the Newtonian space subalgebra, which is given by
 \be
\hat S_{\eta/\kappa}=(\hat x^1)^2 + (\hat x^2)^2 + (\hat x^3)^2  + \frac{\ro}{\kappa}\, \hat x^1 \hat x^2,\qquad [\hat S_{\ro/\kappa} , \hat x^a]=0,
\label{df}
\ee
that can be interpreted as the definition of a ``quantum sphere" in the 3-space, onto which the time operator $\hat x^0$ acts as a dilation:
\be
 [  \hat x^0,\hat S_{\ro/\kappa}]=-\frac 2\kappa\, \hat S_{\eta/\kappa} .
 \label{dg}
\ee
We remark  that the ultimate responsible of the non-vanishing spatial  commutators and of the   existence of the operator $\hat S_{\eta/\kappa}$ is the quantum $\mathfrak{su}(2)\simeq \mathfrak{so}(3)$ subalgebra   generated by the rotation generators $\>J$ (see  (\ref{be}) and (\ref{bi})) appearing  in the $\kappa$-Newtonian deformation, which is a characteristic feature of the  $\kappa$-AdS$_\ka$ deformation and turns out to be preserved under the non-relativistic limit. Note also  that the $\kappa$-Galilei noncommutative spacetime  $\>{G}^{3+1}_\kappa$ with $\ro=0$ is much more degenerate and has the very same abelian spatial sector as the $\kappa$-Minkowski  spacetime (\ref{dc}).

As far as  the $\kappa$-Carrollian Poisson--Hopf algebras deduced in Section~\ref{sec4} are concerned,  there are two different but equivalent ways to obtain their 
associated PHS. On the one hand, since they are coboundary deformations one  can make use of the Sklyanin bracket with classical $r$-matrix    (\ref{cc}), which requires left- and right-invariant vector fields to be computed. On the other hand, one can directly perform the 
ultra-relativistic limit of the $\kappa$-AdS$_\ka$ PHS by following the same approach as in the Newtonian cases, based on the fundamental LBC (which in this case coincides with the coboundary one). Therefore, from the LBC maps 
 (\ref{aj}) and (\ref{cb}) the following transformation on the group parameters are induced
 \be
x^0\to c^{-1}\, x^0,\qquad x^a\to  x^a,\qquad \kappa\to c\,\kappa .
\label{dh}
\ee
If we apply~\eqref{dh} to the $\kappa$-AdS$_\ka$ Poisson brackets (\ref{da}) and (\ref{db}) and compute the limit $c\to 0$, the $\kappa$-Carrollian PHS are obtained, namely
\begin{align}
\begin{split}
\label{di}
&\{x^1,x^0\} =\frac{1}{\kappa}\, \frac{\tanh (\ro x^1)}{\ro \cosh^2(\ro x^2) \cosh^2(\ro x^3)} ,\\
&\{x^2,x^0\} =\frac{1}{\kappa}\,\frac{\tanh (\ro x^2)}{\ro \cosh^2(\ro x^3)} ,\\
&\{x^3,x^0\} =\frac{1}{\kappa}\,\frac{\tanh (\ro x^3)}{\ro},\\
&\{x^a,x^b\} =0.
\end{split}
\end{align} 
We stress that within these spaces the spatial coordinates $x^a$ do Poisson commute, which implies that no ordering ambiguities appear and the $\kappa$-Carrollian noncommutative spacetime is just given by (\ref{di}) by replacing the Poisson brackets by commutators.
Notice that  in this case there is no  quantum  $\mathfrak{su}(2)\simeq \mathfrak{so}(3)$ subalgebra and the rotation generators remain non-deformed (see (\ref{ce}) and (\ref{cf})). This is due to the fact that the  term $J_1\wedge J_2$ in the $r$-matrix  (\ref{ba})  disappears under the ultra-relativistic contraction to the $r$-matrix (\ref{cc}).  Moreover, once more, the proper  $\kappa$-Carroll noncommutative spacetime $\>{C}^{3+1}_\kappa$ with $\ro =0$ coincides with  $\>{M}^{3+1}_\kappa$ (\ref{dc}).


\section{Concluding remarks}
\label{sec6}

While it has been understood for a while that  Planck scale effects might have a non-trivial interplay with curvature effects, their relation to relativistic effects was not investigated before. Here, we provide a first characterization of such interplay by studying $\kappa$-deformed spacetime models with non-vanishing cosmological constant (and correspondingly deformed symmetries) in the non-relativistic $c\to\infty$ and ultra-relativistic $c\to0$ limits. In this way we can categorize the effects that are of purely quantum origin (i.e.~dependent on the quantum deformation parameter $\kappa$),  those that are due to the interplay between the quantum deformation and curvature parameter $\eta$, and those that are also affected by the speed of light parameter $c$.

The first kinds of effects are those that are well-known from the study of the $\kappa$-Minkowski spacetime and $\kappa$-Poincar\'e symmetries. Most prominently, the noncommutativity between the time and space coordinates and a deformation of the algebra of boosts and spatial translations.
Effects that are due to the interplay between $\kappa$ and $\eta$ are the deformation of the algebra of rotations and the noncommutativity between spatial coordinates.
In this paper, we found that the speed of light does not affect those effects that are purely quantum (since they do not appear in  qualitatively different ways in either the non-relativistic limit or the ultra-relativistic limit). In particular, the noncommutativity between time and space coordinates is still present in the two limits, despite the fact that classically one would expect a complete separation between the time and space sectors of spacetime. In this sense, quantum effects are ``stronger'' than relativistic effects. However, $c$ does enter into the picture in a significant way when considering the joint $\kappa$ and $\eta$ effects. For example, the induced modification of the algebra of rotation and the noncommutativity between spatial coordinates are both preserved in the non-relativistic limit, but are lost in the ultra-relativistic limit.

It is worth stressing that the three flat cases with $\ka=\ro =0$ (Minkowski  $\>{M}^{3+1}_\kappa$, Galilei    $\>{G}^{3+1}_\kappa$ and Carroll    $\>{C}^{3+1}_\kappa$) share the same noncommutative spacetime algebra. Therefore it seems rather natural to wonder about the additional symmetry structure that could allow us to distinguish the different features of these three models from a phenomenological perspective. Indeed, the three Hopf algebra structures providing the quantum group invariance of the three spaces are different, and this will be reflected in the properties of their associated curved momentum spaces, as it has  been  recently analysed in~\cite{Flavio}. 
Another framework that might allow to distinguish the three cases is that of  noncommutative  spaces of worldlines associated to a given quantum deformation. This space has  been recently  constructed  for the $\kappa$-Minkowski spacetime  in~\cite{BGH2019worldlinesplb}. In fact, the construction of the  $\kappa$-Galilei and  $\kappa$-Carroll noncommutative spaces of geodesics might give rise to very different geometric structures, which are worth to be studied. Work on this line is currently in progress and will be presented elsewhere.


\section*{Acknowledgements}

 \small
 
This work has been partially supported by Ministerio de Ciencia, Innovaci\'on y Universidades (Spain) under grant MTM2016-79639-P (AEI/FEDER, UE) and by Junta de Castilla y Le\'on (Spain) under grants BU229P18 and BU091G19. The authors acknowledge the contribution of the COST Action CA18108.


\small


\end{document}